\documentclass[superscriptaddress,aps,prb,twocolumn,showpacs,preprintnumbers,amsmath,amssymb]{revtex4}

\usepackage{graphicx}
\usepackage{dcolumn}
\usepackage{bm}
\renewcommand{\Re} {{\rm Re}}
\renewcommand{\Im} {{\rm Im}}
\renewcommand{\Re}{{\rm Re}~}
\renewcommand{\Im}{{\rm Im}~}
\newcommand{\eff}{{\rm eff}}
\newcommand{\es}{{\rm es}}

\def\units#1{{\,\rm{#1}}}

\begin{document}

\preprint{Submitted to Phys.~Rev.~B.(Dec 2008)}

\title{Wide-angle infrared absorber based on negative index plasmonic
  metamaterial}

\author{Yoav Avitzour}
\affiliation{Department of
Physics, The University of Texas at Austin, Austin, Texas 78712}
\author{Yaroslav A. Urzhumov}
\affiliation{COMSOL, Inc., 10850 Wilshire Blvd. \#800, Los Angeles, CA 90024}
\author{Gennady Shvets}
\affiliation{Department of
Physics, The University of Texas at Austin, Austin, Texas 78712}
\email{gena@physics.utexas.edu} 
\date{\today}

\begin{abstract}
  A metamaterials-based approach to making a wide-angle absorber of
  infrared radiation is described. The technique is based on an
  anisotropic Perfectly Impedance Matched Negative Index Material
  (PIMNIM). It is shown analytically that a sub-wavelength in all
  three dimensions PIMNIM enables absorption of close to $100\%$ for
  incidence angles up to $45\deg$ to the normal. A specific
  implementation of such frequency-tunable PIMNIM based on plasmonic
  metamaterials is presented. Applications to infrared imaging and
  coherent thermal sources are described.

\end{abstract}

\maketitle

\section{Introduction}

The emergence of a new field of electromagnetic metamaterials was
brought about by the demand for materials with exotic properties
unattainable in nature. One such property is a negative refractive
index that requires both the effective dielectric permittivity
$\epsilon_{\eff}$ and magnetic permeability $\mu_{\eff}$ to be
negative~\cite{veselago_68,smith_prl00}.  Applications of negative
index metamaterials (NIMs) include ''perfect'' lenses, sub-wavelength
transmission lines and resonators, miniature antennas, among others
~\cite{pendrylens_prl00,Engheta2004,ziolkowski_ieee07}.

The results reported in this paper pertain to two other recently
emerged applications of metamaterials. The first one is
wavelength-selective infrared~\cite{neikirk_ElLett04} and
terahertz~\cite{padilla_optexp_08,TaoH08_PRB} detection, important for
thermal imaging, night vision, and nondestructive
detection. Wide-angle power absorption efficiency is desirable for
miniaturizing photodetectors or microbolometers down to the wavelength
size. The second application is the development of coherent thermal
emitters~\cite{fleming_nature_02,chan_pre_06,greffet_nature02,hamann_optexp07}
for spectroscopic and thermophotovoltaic (TPV)
~\cite{coutts_renewable99,laroche_apl06} applications. By virtue of
the Kirchhoff's Law, emissivity of a thermal emitter approaches the
blackbody limit only if the absorptivity approaches unity.  Moreover,
wavelength-selective radiators can dramatically improve the efficiency
of current generation in a TPV
cell~\cite{coutts_renewable99,laroche_apl06} if their emission
spectrum is matched to the bandgap of the TPV converter. For example,
a typical TPV converter, GaSb, has the bandgap of $E_G =
0.7\units{eV}$ that would be ideally suited to a wavelength-selective
radiator operating in near-infrared (around $\lambda = 1.7 \units{\mu
  m}$).  For spectroscopic applications, arrays of wavelength-sized
narrow-band coherent thermal emitters based on a metamaterial unit
cell can be designed. The emitted radiation of each detector is
focusable to wavelength-sized spots using the far-field by large
numerical aperture optics.  With these applications in mind, we
describe in this paper a ultra-thin metamaterials-based wide-angle
absorber of near-infrared radiation. The design of this perfect
absorber is inspired by a Perfectly Impedance Matched Negative Index
Metamaterial (PIMNIM) constructed from plasmonic wires.

\section{Theoretical Background}
\begin{figure}[ht]
\begin{center}
\includegraphics[width=0.9\linewidth]{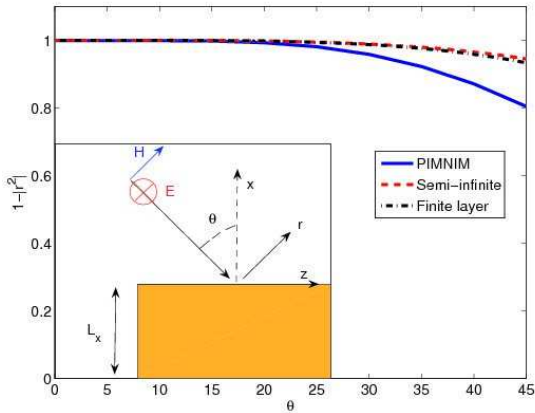}
\caption{(Color online) Angular dependence of the absorption coefficient for the
  idealized structure with $\mu_{zz}=\epsilon_{yy}=-1+i$,
  $\mu_{xx}=1$, and $L_x = \infty$ (dashed line), $L_x = 200\units{nm}$
  (dashed-dotted line), barely distinguishable from the $L_x=\infty$
  line. Solid line: same for the specific implementation of the
  metamaterial absorber shown in Fig.~\ref{fig:schematic}. A schematic
  of the idealized wide-angle metamaterial absorber is given in the
  inset, S-polarized incident wave is assumed.}
\label{fig:absorber_schematic}
\end{center}
\end{figure}

As a background, we consider a simple problem of radiation absorption
by a semi-infinite slab of a lossy metamaterial with engineered
dielectric permittivity and magnetic permeability tensors
$\bar{\bar{\epsilon}}$ and $\bar{\bar{\mu}}$. Radiation is assumed to
be incident in the $x-z$ plane at an angle $\theta$ with respect to
the vacuum-material interface normal $\vec{e}_x$.  We further assume
$s-$polarization of the incident wave as shown in the inset of
Fig.~\ref{fig:absorber_schematic}, i.e., the only non-vanishing field
components are $E_y, H_x, H_z$. Assuming that both
$\bar{\bar{\epsilon}}$ and $\bar{\bar{\mu}}$ are diagonal tensors,
their only relevant components are $\epsilon_{yy}$, $\mu_{xx}$, and
$\mu_{zz}$. For our semi-infinite slab (assuming that the
metamaterial's thickness $L_x$ is sufficient to absorb all transmitted
radiation), absorptivity $A$ is limited only by reflection .  A
straightforward calculation yields the reflection $r$ and absorption
$A = 1 - |r|^2$ coefficients:
\begin{equation}\label{eq:absorptivity}
    A = 1 - \left| \frac{\cos{\theta} -
    \sqrt{\epsilon_{yy}/\mu_{zz} - \sin^2{\theta}/(\mu_{zz}
    \mu_{xx})}}{\cos{\theta} +
    \sqrt{\epsilon_{yy}/\mu_{zz} - \sin^2{\theta}/(\mu_{zz}
    \mu_{xx})}} \right|^2,
\end{equation}
where all material parameters are, in general,
wavelength-dependent.

Studying Eq.~(\ref{eq:absorptivity}), we note that total absorption
($A(\lambda_0) \equiv 1$) at a specific wavelength $\lambda_0$ is
achievable at normal incidence ($\theta=0$) if the absorber material's
impedance $\eta = \sqrt{\mu_{zz}/\epsilon_{yy}}$ is matched to that of
vacuum, i.e., when $\mu_{zz}(\lambda_0) =
\epsilon_{yy}(\lambda_0)$. Using impedance matching to achieve total
absorption is a very well known approach in microwave
engineering~\cite{shin_ieee93,padilla_absorber_08}. More surprising is
another prediction of Eq.~(\ref{eq:absorptivity}): that nearly total
absorption can be achieved over a very broad range of angles. Assuming
that $\mu_{xx}(\lambda_0) = 1$ and $|\mu_{zz}|=|\epsilon_{yy}| \gg 1$,
we find that $A(\lambda_0) \approx 1 - \tan^4{\theta/2}$. For a
$90\deg$ full-angle ($\theta_{max}=\pi/4$) we find that $0.97 < A < 1$
for $0 < \theta < \theta_{max}$. Even more remarkably, a similarly
high broad-angle absorption is predicted even for much modest values
of $\mu_{zz}$ and $\epsilon_{yy}$. For example, if the absorbing
medium is a Negative Index Material
(NIM)~\cite{veselago_68,smith_prl00} with
$\epsilon_{yy}=\mu_{zz}=-1+i$, we find that $0.94 < A < 1$. The
angular dependence $A(\theta)$ for such perfectly impedance matched
negative index material (PIMNIM) is shown in
Fig.~\ref{fig:absorber_schematic} for the semi-infinite metamaterial
slab, $L_x=\infty$, as well as for the more general case of a finite
slab, with $L_x \approx \lambda_0/7$. The two curves are barely
distinguishable because of the high metamaterial loss.

The main implication of these results is that a sub-wavelength slab of
{\it almost any} impedance-matched metamaterial acts as a wide-angle
absorber. The challenge now is to design such a metamaterial,
especially in the technologically important infrared part of the
spectrum. With this goal in mind, the rest of this paper is organized
as follows. In section \ref{sec:design} we present one such possible
metamaterial design based on two layers of alternating long and short
plasmonic nanoantennas (Fig.~\ref{fig:schematic}), and demonstrate
that the unit cell of this PIMNIM can be made highly sub-wavelength in
near-infrared. The sub-wavelength requirement is critical for
achieving wide-angle absorption: recent numerical
simulations~\cite{padilla_absorber_08} have found that $A(\theta)$
rapidly drops with $\theta$ when the unit cell is too large.
Then, in section \ref{sec:simulation}, we demonstrate that a single
layer of PIMNIM enables optical absorption $0.7< A(\theta) < 0.9$ for
a $90\deg$ full-angle scan, and that its small-angle absorption
coefficient exactly agrees with the prediction of
Eq.~(\ref{eq:absorptivity}).  Finally, in section \ref{sec:discussion}
we show that extremely high optical intensity enhancements (at least
$\times 10^3$) can be produced at the absorption peak.

\section{PIMNIM Design}
\label{sec:design}
\begin{figure}[ht]
\begin{center}
\includegraphics[width=0.9\linewidth]{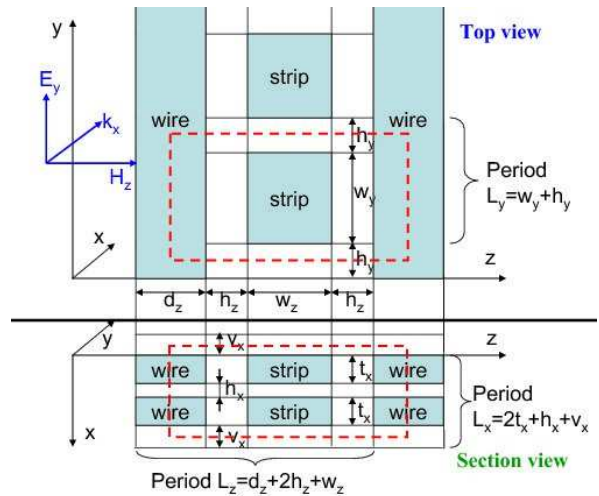}
\caption{(Color online) Schematic of the PIMNIM structure. Unit cell
  for electromagnetic and electrostatic simulations is inside the
  dashed rectangles. } \label{fig:schematic}
\end{center}
\end{figure}

The geometry of the unit cell of the proposed PIMNIM is shown in
Fig.~\ref{fig:schematic}. Each unit cell consists of two parallel
layers separated by the distance $h_x$. Each layer consists of a
cut-wire of width $w_z$ and length $w_y$ surrounded by the continuous
in the $y-$direction wires of width $d_z$. To enable planar
fabrication, the heights $t_x$ of the cut and continuous wires are
assumed to be the same. Metal wires are assumed to be embedded inside
a dielectric with $\epsilon_d = 2.25$. An s-polarized electromagnetic
wave $\vec{E} \parallel \vec{e}_y$ excites the electric and magnetic
responses of the unit cell. The magnetic response is caused by
counter-propagating currents flowing through the adjacent
cut-wires. Both cut and continuous wires contribute to the electric
response. A similar structure has been analyzed in the microwave part
of the spectrum~\cite{kafesaki_fishnet07}, where it is not
sub-wavelength. {\it Plasmonic effects}, i.e. taking into account that
metals have a finite dielectric permittivity $\epsilon_m \equiv
\epsilon_m^{\prime} + i \epsilon_m^{\prime \prime}$ with
$\epsilon_m^{\prime} < 0$, are {\it necessary} to miniaturize the unit
cell.

Electromagnetic resonances in effective permittivity $\epsilon_\eff$
and magnetic permeability $\mu_\eff$ of plasmonic composites are
unambiguously related
~\cite{urzh_shvets_ssc08,urzh_nordlander07,shvets_urzh_prl04} to the
electrostatic surface plasmon resonances of the appropriate symmetry
(electric dipole and magnetic dipole, correspondingly).  Therefore,
the first step in investigating the suitability of a particular unit
cell geometry for a plasmonic NIM is to identify the frequencies of
its electrostatic resonances. Specifically, magnetic activity
(including negative index behavior) has been shown to
exist~\cite{urzh_shvets_ssc08} only for $\lambda > \lambda_{\es}$,
where $\lambda_{\es} \equiv 2\pi c/\omega_{\es}$ is the vacuum
wavelength corresponding to the frequency of the electrostatic
resonance responsible for the magnetic activity.  Such
magnetically-active (MA) resonance of the PIMNIM structure has been
calculated by solving the Poisson's equation $\vec{\nabla} \cdot
\left( \epsilon \vec{\nabla} \phi \right)=0$ for the electrostatic
potential $\phi$, where $\epsilon$ is a piecewise constant function
equal to $\epsilon_m$ inside the metal wires and $\epsilon_d$
outside. Poisson's equation can be solved as a generalized eigenvalue
equation~\cite{bergman92,stockman_bergman01,shvets_urzh_prl04} to
yield the values of $\epsilon_m$ corresponding to the electrostatic
resonances.

\begin{figure}[ht]
\begin{center}
\includegraphics[width=0.75\linewidth]{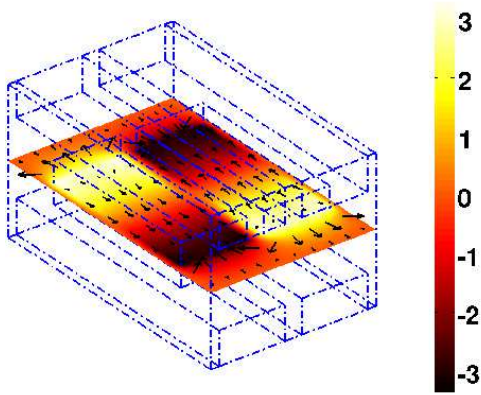}
\caption{(Color online) Electrostatic magnetically-active plasmon
resonance of the PIMNIM structure shown in
Fig.~\ref{fig:schematic} with the scalable geometric proportions
corresponding to (in nm): $L_y=320$, $w_y=200$, $t_x=80$,
$w_z=50$, $d_z=80$, $h_x=20$, $v_x=10$, $h_z=20$.
The resonance occurs at $\epsilon_m/\epsilon_d=-26.8$. Arrows:
electric field; color: electrostatic potential.}
\label{fig:magn_es_eig_set2}
\end{center}
\end{figure}

Potential distribution of the lowest-frequency MA resonance is shown
in Fig.~\ref{fig:magn_es_eig_set2}. Its magnetic nature can be deduced
from the electric field loops formed between the two cut-wires. The
above resonance occurs when the dielectric contrast between metal
wires and the dielectric matrix is
$\epsilon_m/\epsilon_d=-26.8$. Because the resonant dielectric
contrast is determined from electrostatic calculations that contains
no spatial scale, this value is determined by the {\it geometric
  shape} of the unit cell and its inclusions but not by the overall
scale set by either of the periods $L_{x,y,z}$. Assuming gold as the
plasmonic component and silica with $\epsilon_d=2.25$ as an embedding
dielectric, the electrostatic resonance occurs at
$\lambda_{\es}=1.2\units{\mu m}$, thereby setting the lower limit on
the wavelength at which strong magnetic response is expected. The
next, electrically active (EA) resonance corresponds to
$\epsilon_m/\epsilon_d=-19$, is considerably blue-shifted from the MA
resonance. Therefore, the PIMNIM structure depicted in
Fig.~\ref{fig:magn_es_eig_set2} is promising as a NIM that has widely
separated electric and magnetic resonances. Thus, its electromagnetic
characteristics strongly resemble those of the original microwave
NIMs~\cite{smith_prl00}. The important difference is that the proposed
structure operates at the near-infrared frequencies and exploits
plasmonic resonances to achieve sub-$\lambda_d$ cell size, where
$\lambda_d$ is the wavelength inside the substrate.

It is instructive to compare the resonant frequency of the MA
resonance of the PIMNIM structure with that of the traditional Double
Fishnet (DF)~\cite{brueck_prl05,wegener_science06} structure. DF can
be obtained from the PIMNIM by extending the cut-wires in
$z$-direction until they merge with continuous wires running along the
$y$-axis, thereby forming the second, orthogonal set of continuous
wires in the $z$-direction. For example, the DF structure thus
obtained from the PIMNIM structure shown in
Fig.~\ref{fig:magn_es_eig_set2} is periodic in the $y-z$ plane with
the periods of $L_z = 170\units{nm}$ and $L_y = 320\units{nm}$
containing two sets of intersecting metal strips in $y$ and $z$
direction, with the corresponding widths of $d_z = 80\units{nm}$ and
$w_y = 200\units{nm}$. The MA plasmonic resonance is found at
$\epsilon_m/\epsilon_d=-8.3$ which corresponds to
$\lambda_{\es}=0.73\units{\mu m}$ for gold. Of course, the physical
sizes can be scaled from the above dimensions by an arbitrary factor
because there is no physical scale in electrostatics. This simulation
illustrates the challenge of making a strongly sub-$\lambda_d$ DF
structure. For the unit cell to be sub-$\lambda_d$, the operational
wavelength must be reasonably close to $\lambda_{\es}$. Therefore,
making a unit cell with the largest dimension of $\lambda/4$ requires
$L_y < 200\units{nm}$.  Such small DF structures have never been
fabricated to date, which explains why sub-$\lambda_d$ DF-based NIM
have never been produced.

\section{Simulation Results}
\label{sec:simulation}

\begin{figure}[ht]
\begin{center}
\includegraphics[width=0.9\columnwidth]{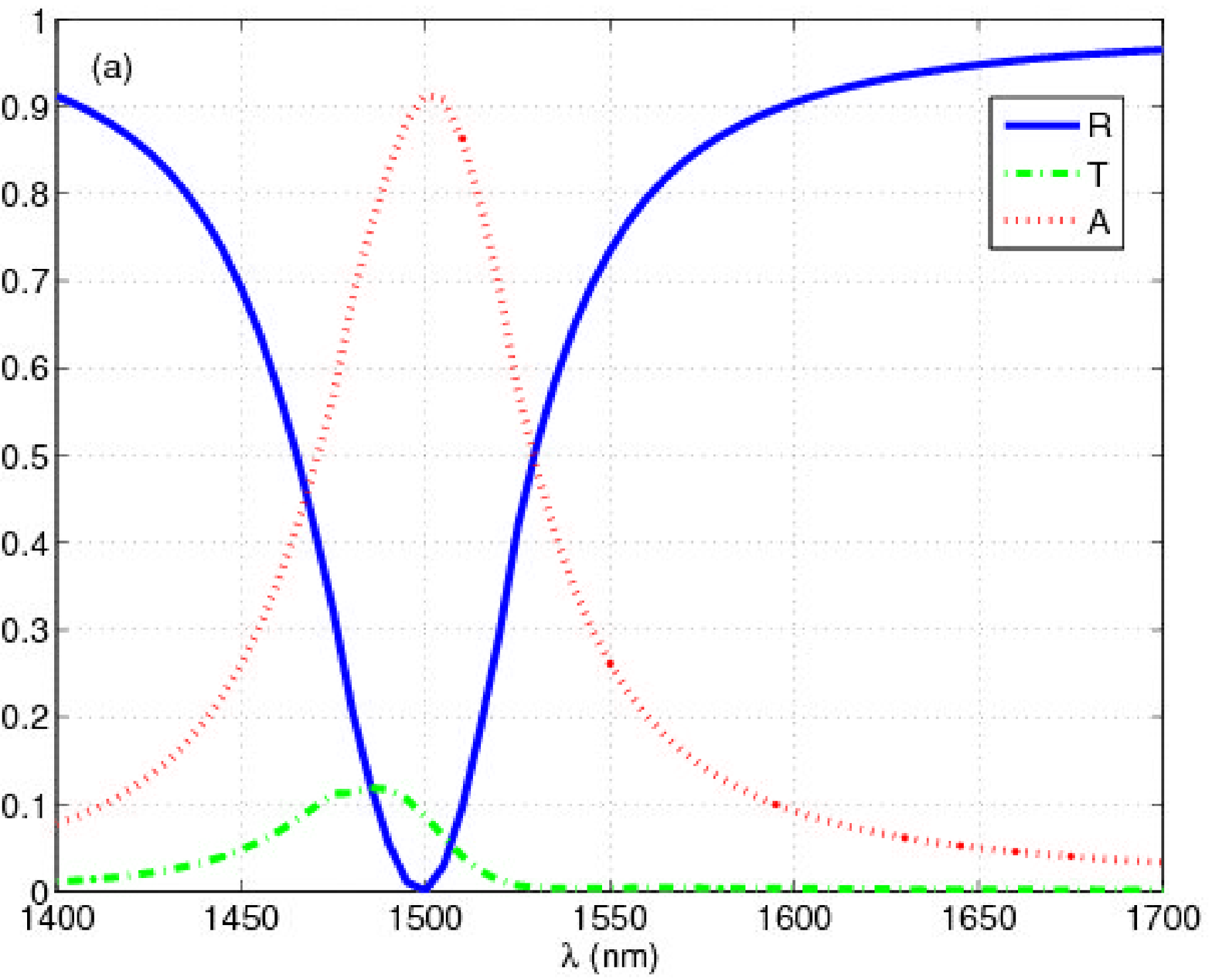}
\includegraphics[width=0.9\columnwidth]{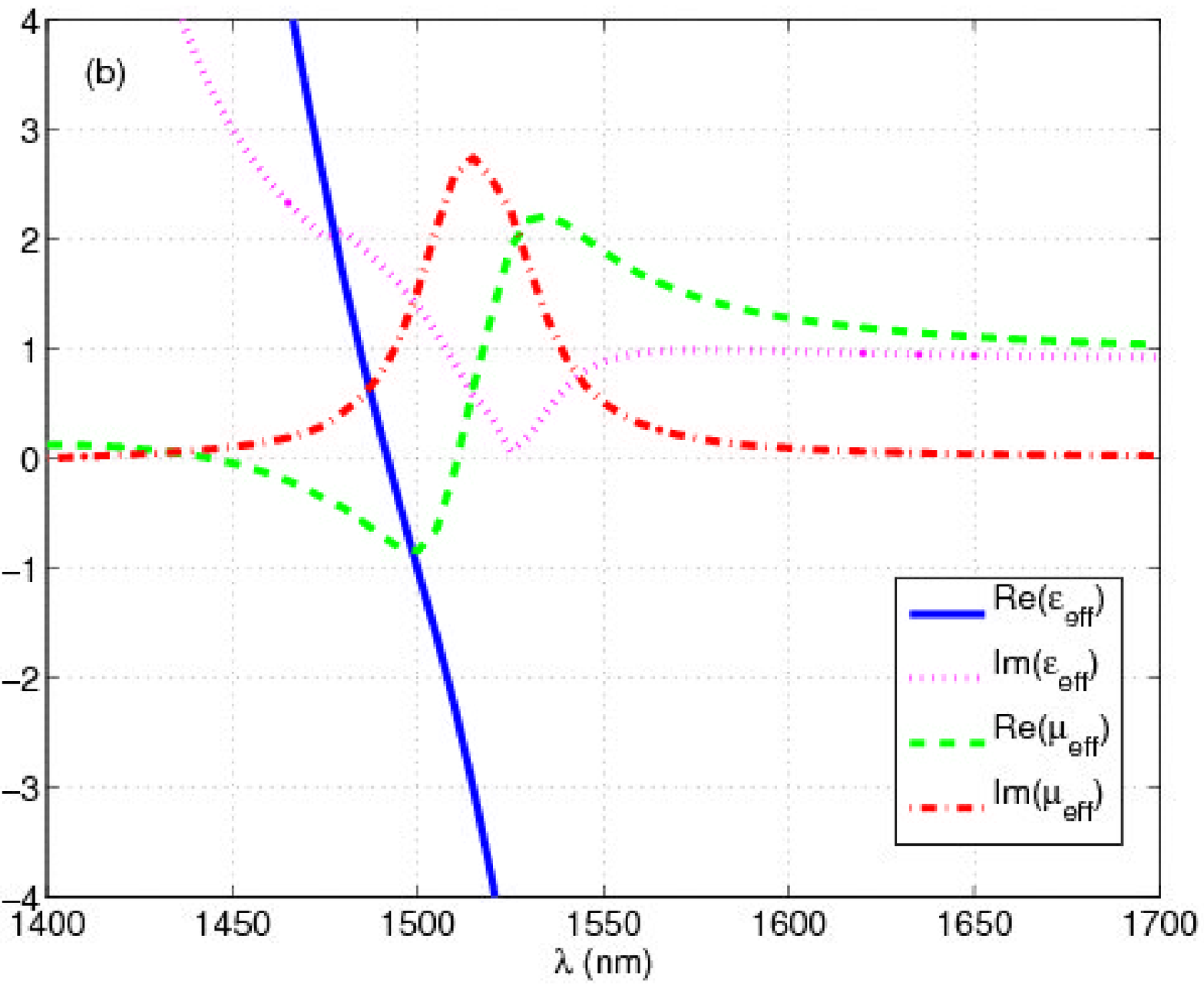}
\caption{(Color online) (a) Reflectance, transmittance and absorbance
  at normal incidence of a single PIMNIM layer. (b) Extracted
  effective dielectric permittivity $\epsilon_{\eff}$ and magnetic
  permeability $\mu_{\eff}$ of the PIMNIM structure. Reflectance
  vanishes at $\lambda=1.5\units{\mu m}$ because
  $\epsilon_{\eff}=\mu_{\eff}$.  Parameters (in nm): $L_y=326$,
  $w_y=208$, $t_x=80$, $w_z=46$, $d_z=78$, $h_x=21$,
  $h_z=20$. } \label{fig:RTA_set4}
\end{center}
\end{figure}

Full electromagnetic finite elements frequency domain (FEFD)
simulations of light transmission/reflection through the PIMNIM
structure shown in Fig.~\ref{fig:schematic} (approximate physical
parameters indicated in Fig.~\ref{fig:magn_es_eig_set2} and further
optimized to achieve perfect impedance matching) were carried out
using the \textsc{COMSOL}~\cite{comsol06} commercial package. Drude
model for $\epsilon_m(\omega)$ of gold, with $\omega_p=1.367\times
10^{16}\units{rad/sec}$ and $\gamma=4\times 10^{13}\units{rad/sec}$
(taken from \cite{ordal_apopt83}), and $\epsilon_d=2.25$ were used in
the simulations. Reflection minimization was carried out using
standard non-linear multivariable optimization to fine-tune the PIMNIM
parameters and adjust the perfect-matching wavelength $\lambda_0$.
The normal incidence ($\theta=0$) results are shown in
Fig.~\ref{fig:RTA_set4}(a): zero reflection is achieved (by design) at
$\lambda_0=1.5\units{\mu m}$, with the single PIMNIM layer absorption
coefficient $A \approx 0.9$ (or $A \approx 0.99$ for two layers)

\begin{figure}
\begin{center}
\includegraphics[width=0.9\columnwidth]{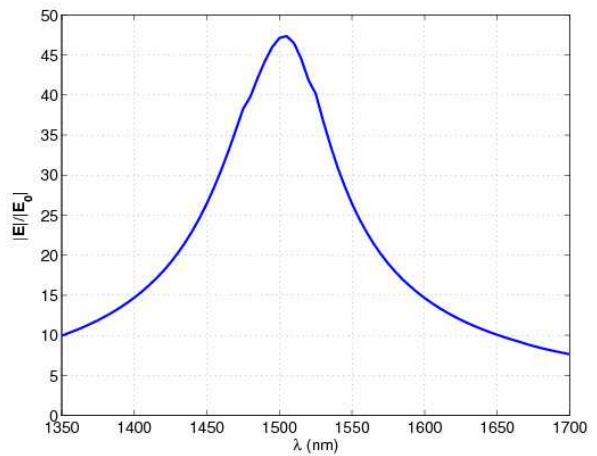}
\caption{\label{fig:field_enhancement_range} Maximum field
  enhancement, $|\bm{E}|/|E_0|$, in the PIMNIM vs. wavelength. It is noted
  that the enhancement is maximized at the absorption peak, and has no
  structure corresponding to the electric resonance at
  $\lambda=1380\units{nm}$.}
\end{center}
\end{figure}

\begin{figure}[ht]
\begin{center}
\includegraphics[width=0.9\columnwidth]{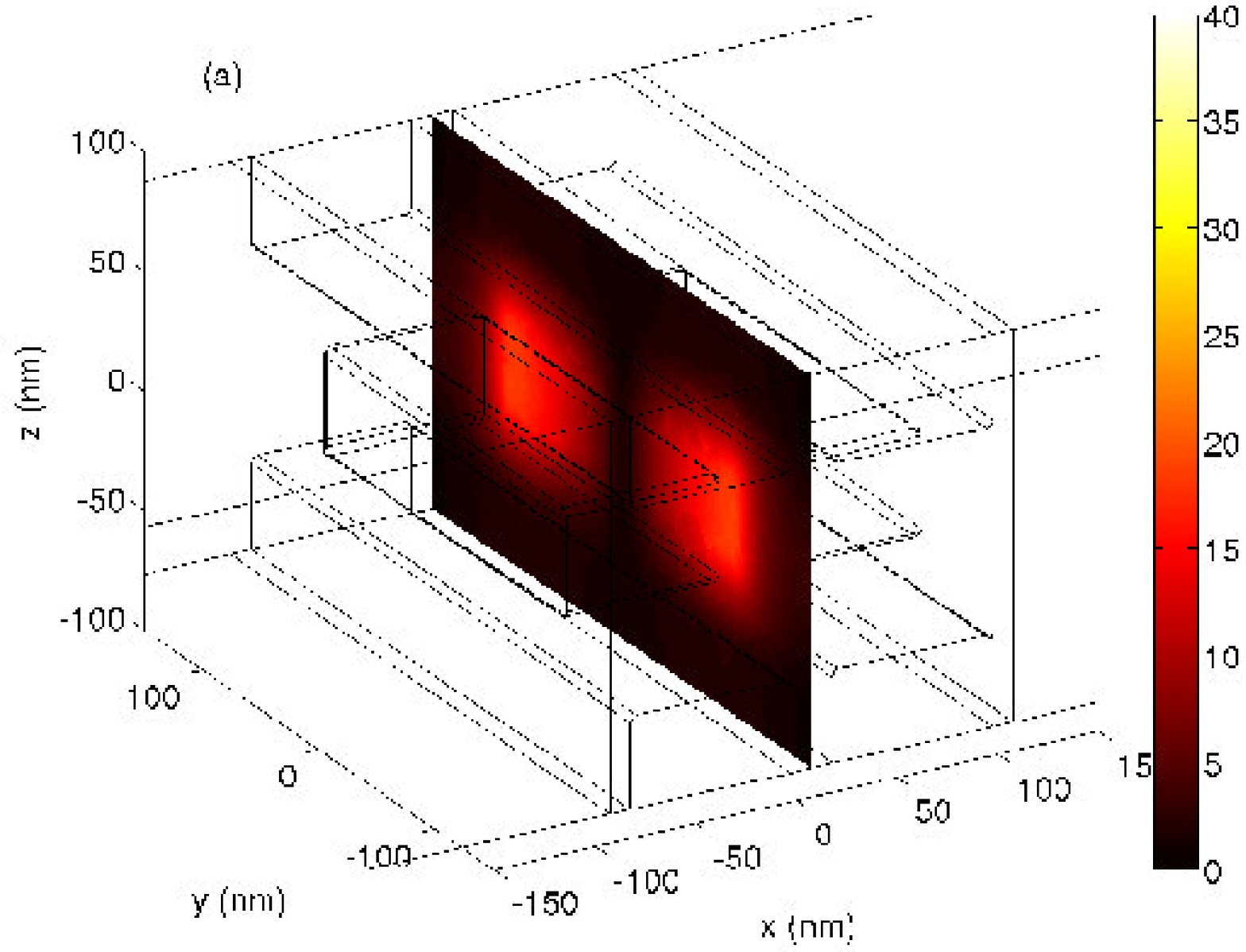}
\includegraphics[width=0.9\columnwidth]{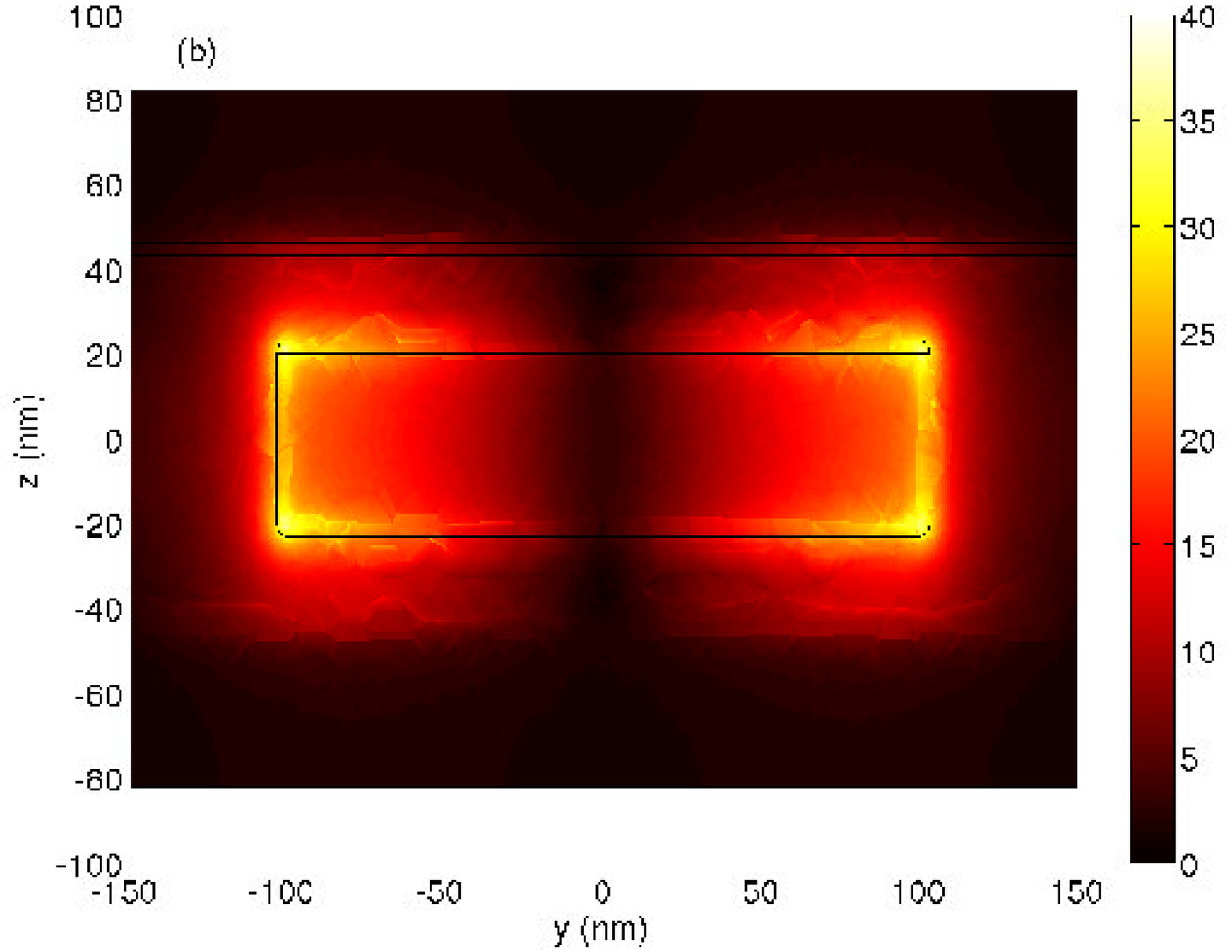}
\caption{\label{fig:field_enhancement}(Color online) Field
  enhancement, $|\bm{E}|/|E_0|$ in the PIMNIM. Field is plotted on $y-z$
  slices at (a) $x=0$ and (b) $x=-8\units{nm}$. Although the largest
  enhancement is near the edges and corners, a relatively large area
  experiences significant enhancement.} 
\end{center}
\end{figure}
Effective parameters $\epsilon_{yy}\equiv \epsilon_{\eff}(\lambda)$
and $\mu_{zz}\equiv \mu_{\eff}(\lambda)$ of the PIMNIM structure were
extracted from the complex transmission and reflection coefficients
through a single PIMNIM layer using the standard
procedure~\cite{smith_prb02}. Because the negative index band is the
lowest (in frequency) transmission band of the PIMNIM, there is no
ambiguity in determining $\epsilon_{\eff}$ and $\mu_{\eff}$. At the
impedance matching point $\lambda_0=1.5\units{\mu m}$ it is found that
$\epsilon_{\eff}=\mu_{\eff}=-0.85+1.4i$. Although the so-called
figure-of-merit (FOM=$\Re n/\Im n$) is less than unity, achieving high
FOM is not necessary (or even desirable) for accomplishing total
wide-angle absorption.  More important is ensuring that the unit cell
of the PIMNIM is very sub-wavelength: $L_{x,y,z} \ll \lambda_d$. This
is indeed accomplished by the present PIMNIM design: $n_d
L_y/\lambda_0 \approx 0.3$.

The peak absorption, $A_{\rm max}(\theta) \equiv 1 - |r|^2(\theta)$,
of the PIMNIM is plotted in Fig.~\ref{fig:absorber_schematic}. For
small $\theta$ there is an excellent agreement between $A_{\rm max}$
and Eq.~(\ref{eq:absorptivity}) indicating that the PIMNIM is
accurately described as an effective medium. Even for $\theta=\pm
45\deg$ absorptivity remains at about $70\%$. This implies that, in a
narrow spectral interval around $\lambda_0$, emissivity of a heated
PIMNIM is close to that of a blackbody over a broad range of angles,
which would enable to manufacture wavelength-size coherent thermal
emitters.

\section{Field Enhancement}
\label{sec:discussion}
Physically, the absorption peak corresponds to strong electric field
enhancement inside the PIMNIM structure. For example, for the PIMNIM
described by Fig.~\ref{fig:RTA_set4}, it is found that the maximum
intensity enhancement near the corners and edges of the cut-wire
exceeds $|E_{\rm max}|^2/|E_0|^2 > 2000$, where $E_0$ is the amplitude
of the incident wave, as can be seen in
Fig. \ref{fig:field_enhancement_range}. We note that calculating the
maximum field in the simulation domain can be imprecise, because mesh
irregularities can lead to mesh-dependent spikes at the corners and
edges of the structure. In order to calculate the field enhancement
while avoiding such numerical artifacts, the results presented in
Fig. \ref{fig:field_enhancement_range} were calculated by taking the
maximum of $|\bm{E}|$ interpolated on a rectangular grid of $1\times
1\times 1\units{nm}$ and then smoothed by nearest-neighbor
smoothing. This method was found to be robust under variations in mesh
structure and density. Interestingly, the field enhancement is
maximized at the absorption peak, at $\lambda=1.5\units{\mu m}$, yet
shows no structure near the electric resonance, at
$\lambda=1.38\units{\mu m}$. This can be explained by the very strong
reflection at the electric resonance, which prevents the incident
fields from penetrating into structure.

In addition to the enhancement near the corners and edges of the
structure, fairly large field enhancement exists in a considerable
volume between the cut wires. Fig.~\ref{fig:field_enhancement}
presents $|\bm{E}|$ at two $y-z$ cross-sections, for different values
of $x$. In Fig.~\ref{fig:field_enhancement}(a) the magnitude of the
electric field $|\bm{E}|$ is plotted at $x=0$ (in the middle of the
PIMNIM).  Intensity enhancement of $|\bm{E}|^2/|E_0|^2 \approx 200$ is
apparent in about two thirds of the region between the front and the
back cut-wires.  Fig.~\ref{fig:field_enhancement}(b) presents the
field closer the front cut-wire, at $x=-8\units{nm}$. An average
intensity enhancement of $|\bm{E}|^2/|E_0|^2 \approx 400$ is apparent
in roughly the same area. If photon-counting detectors are integrated
into the PIMNIM structure, then such intensity enhancement translates
into proportionally enhanced absorption efficiency.  The consequence
of the large field enhancement is the desirable ultra-thin dimension
of the absorber, in contrast to earlier
calculations~\cite{Yannopapas06} that demonstrated that thicker (about
one wavelength) perfect absorbers can be developed using plasmonic
spheres.

\section{Conclusions}
In conclusion, we have demonstrated that an impedance-matched negative
index metamaterial can act as a wavelength-selective wide-angle
absorber of infra-red radiation. A specific implementation of such
frequency-tunable PIMNIM based on plasmonic metamaterials is
presented. Applications of the PIMNIM include infrared imaging and
coherent thermal sources. This work was supported by the ARO MURI
W911NF-04-01-0203, AFOSR MURI FA9550-06-1-0279, and the NSF NIRT grant
No. 0709323.

\bibliography{PIMNIM_bib}
\end{document}